\documentclass[pre,twocolumn,a4paper,showpacs]{revtex4-1}
\usepackage{amssymb}
\usepackage{wasysym}
\usepackage{graphicx}
\usepackage{epsfig}
\usepackage{subfigure}
\begin{document}

\title{Exploring NK Fitness Landscapes Using  Imitative Learning}

\author{Jos\'e F. Fontanari}

\affiliation{Instituto de F\'{\i}sica de S\~ao Carlos,
  Universidade de S\~ao Paulo,
  Caixa Postal 369, 13560-970 S\~ao Carlos, S\~ao Paulo, Brazil}

\begin{abstract}
The idea that a group of cooperating agents can solve problems more efficiently than when those agents work independently is hardly controversial, despite our obliviousness of the conditions that make cooperation a successful problem solving strategy. 
Here we investigate the performance of a group of agents in locating the global maxima of  NK fitness landscapes with varying degrees of ruggedness. Cooperation is taken into account through imitative learning and the
broadcasting of  messages informing on the fitness of each agent.
We find a trade-off between the  group size and the frequency of  imitation: for rugged landscapes, too much imitation or  too large a group yield a performance  poorer than that of independent agents. By decreasing the diversity of the group, imitative learning  may lead to duplication of work and hence to a decrease of  its effective size. However, when the parameters are set to optimal values  the cooperative group substantially outperforms the independent agents.
\end{abstract}

\maketitle

\section{Introduction}\label{sec:intro}

The benefits of cooperative work were  explored by nature well before the advent of the human species as attested by the collective structures built by slime molds and social insects \cite{Wilson_75}. In the human context, the socio-cognitive niche hypothesis purports that hominin evolution relies so heavily on social elements  to  the point of viewing a band of hunter--gatherers as  a  `group-level predator' \cite{Whiten_12}. 
Whether physical or social, those structures and organizations  may be thought of as the organisms' solutions to the problems that endanger their existence (see, e.g., \cite{Bloom_01})  and have motivated the  introduction of the concept of social intelligence in the scientific milieu \cite{Nehaniv_07}. 
In the computer science circle, that concept prompted  the proposal of optimization heuristics based on social interaction, such as the popular particle swarm optimization algorithm \cite{Kennedy_99,Bonabeau_99} and the perhaps lesser-known adaptive culture heuristic \cite{Kennedy_98,Fontanari_10}.

Despite the prevalence of the notion that cooperation can aid a group of agents to solve problems more efficiently than if those agents worked in isolation, and of the success of the social interaction heuristics in producing near optimal solutions to a variety of combinatorial optimization problems, we know little about the conditions  that make cooperative work more efficient than independent work.  In particular, we note that since cooperation (or communication, in general) reduces the diversity or heterogeneity of the group of agents, it  may actually end up reducing the efficiency of group work   \cite{Hong_01,Page_07}. Efficiency here means that the time to solve a problem scales superlinearly with the number of individuals or resources employed in the task.

In this contribution we study the performance of a group of cooperative agents following a first-principle research strategy to study  cooperative problem solving set forth by Huberman in the 1990s that consists of tackling easy tasks, endowing the agents with simple random trial-and-test search tools, and using plain protocols of collaboration \cite{Huberman_90,Clearwater_91}. Here the task is to find the global maxima of three realizations of the NK-fitness landscape characterized by different degrees of ruggedness (i.e., values of the parameter $K$). We use a group of M agents which, in addition to the ability to perform individual trial-and-test searches, can imitate a model agent -- the best performing agent at the trial -- with a probability $p$.  Hence our model exhibits the two critical ingredients of a collective brain according to Bloom: imitative learning and a dynamic hierarchy among the agents \cite{Bloom_01}. The model exhibits also the key feature of distributed cooperative problem solving systems, namely, the exchange of messages   between agents  informing each other on their partial success
(i.e., their fitness at the current trial)  towards the completion of the goal \cite{Huberman_90}.

We find that the presence of local maxima  in the fitness landscape introduces a complex trade-off between the computational cost to solve the task  and the group size $M$. In particular, for a fixed frequency of imitation $p$, there is an optimal value of $M$ at which  the computational cost is minimized. This finding leads to the conjecture that 
the efficacy of imitative learning could be a factor determinant of the group size of social animals \cite{Fontanari_14}.

Our study departs from the vast literature on cooperation that followed Robert Axelrod's 1984  seminal book The Evolution of Cooperation  \cite{Axelrod_84} since in that game theoretical approach it is usually assumed a priori that  mutual cooperation is the most rewarding strategy to the individuals. On the other hand, here we consider a problem solving scenario  and a specific cooperation mechanism (imitation) aiming at determining in which conditions cooperation is more efficient than  independent work.

The rest of the paper is organized as follows. In Section \ref{sec:NK} we offer a brief overview of the NK model of  
rugged fitness landscapes.  In Section \ref{sec:Imi}  we present a detailed account of the imitative learning search strategy and in  Section \ref{sec:res} we study its performance on the task of finding the global maxima of three realizations of  the NK  landscape with $N=12$ and
ruggedness parameters
$K=0$, $K=2$ and $K=4$. The rather small size of the solution space ($2^{12} = 4096$ binary strings of length $12$)  allows the full exploration of the space of parameters and, in particular, the study of the regime where the group size is much greater than the solution space size.  Finally, Section \ref{sec:conc} is reserved to our concluding remarks.

\section{NK Model of Rugged Fitness Landscapes}\label{sec:NK}

The NK model of rugged fitness landscapes introduced by Kauffman \cite{Kauffman_87,Kauffman_89} offers the ideal framework to test the potential of imitative learning in solving optimization problems of varied degrees of difficulty,  since the ruggedness of the landscape can be tuned by changing the two integer parameters -- $N$ and $K$ --  of the model. Roughly speaking, the parameter $N$ determines the size of the solution space whereas the value of $K =0, \ldots, N-1$ influences the number of local maxima and minima on the landscape.

The solution space consists of the $2^N$ distinct binary strings of length $N$, $\mathbf{x} = \left ( x_1, x_2, \ldots,x_N \right )$ with
$x_i = 0,1$. To each string we associate a fitness value $\Phi \left ( \mathbf{x}  \right ) $ which  is an average  of the contributions from each 
component $i$ in the string, i.e.,
\begin{equation}
\Phi \left ( \mathbf{x}  \right ) = \frac{1}{N} \sum_{i=1}^N \phi_i \left (  \mathbf{x}  \right ) ,
\end{equation}
where $ \phi_i$ is the contribution of component $i$ to the  fitness of string $ \mathbf{x} $. It is assumed that $ \phi_i$ depends on the state
$x_i$  as well as on the states of the $K$ right neighbors of $i$, i.e., $\phi_i = \phi_i \left ( x_i, x_{i+1}, \ldots, x_{i+K} \right )$ with the arithmetic in the subscripts done modulo $N$. In addition, the functions $\phi_i$ are $N$ distinct real-valued functions on $\left \{ 0,1 \right \}^{K+1}$. As usual,  we assign to each $ \phi_i$ a uniformly distributed random number  in the unit interval \cite{Kauffman_87,Kauffman_89}. Hence $\Phi \in \left ( 0, 1 \right )$ has a unique global maximum. 

For $K=0$ there are no local maxima and the sole maximum of $\Phi$ is easily located by picking for each component $i$ the state $x_i = 0$ if  $\phi_i \left ( 0 \right ) >  \phi_i \left ( 1 \right )$ or the state  $x_i = 1$, otherwise. For $K>0$, finding the global maximum of the NK model is a NP-complete problem \cite{Solow_00}, which essentially means that the time required to solve the problem using any currently known deterministic algorithm increases exponentially fast  as the size $N$ of the problem  grows \cite{Garey_79}.
The increase of the parameter $K$ from $0$ to $N-1$  decreases the correlation between the fitness of neighboring configurations 
(i.e., configurations that differ at a single component) in the solution space and for $K=N-1$ those fitness values are  uncorrelated
so the NK model reduces to the Random Energy model \cite{Derrida_81,David_09}.
%
\begin{figure}
\centering
\includegraphics[width=0.48\textwidth]{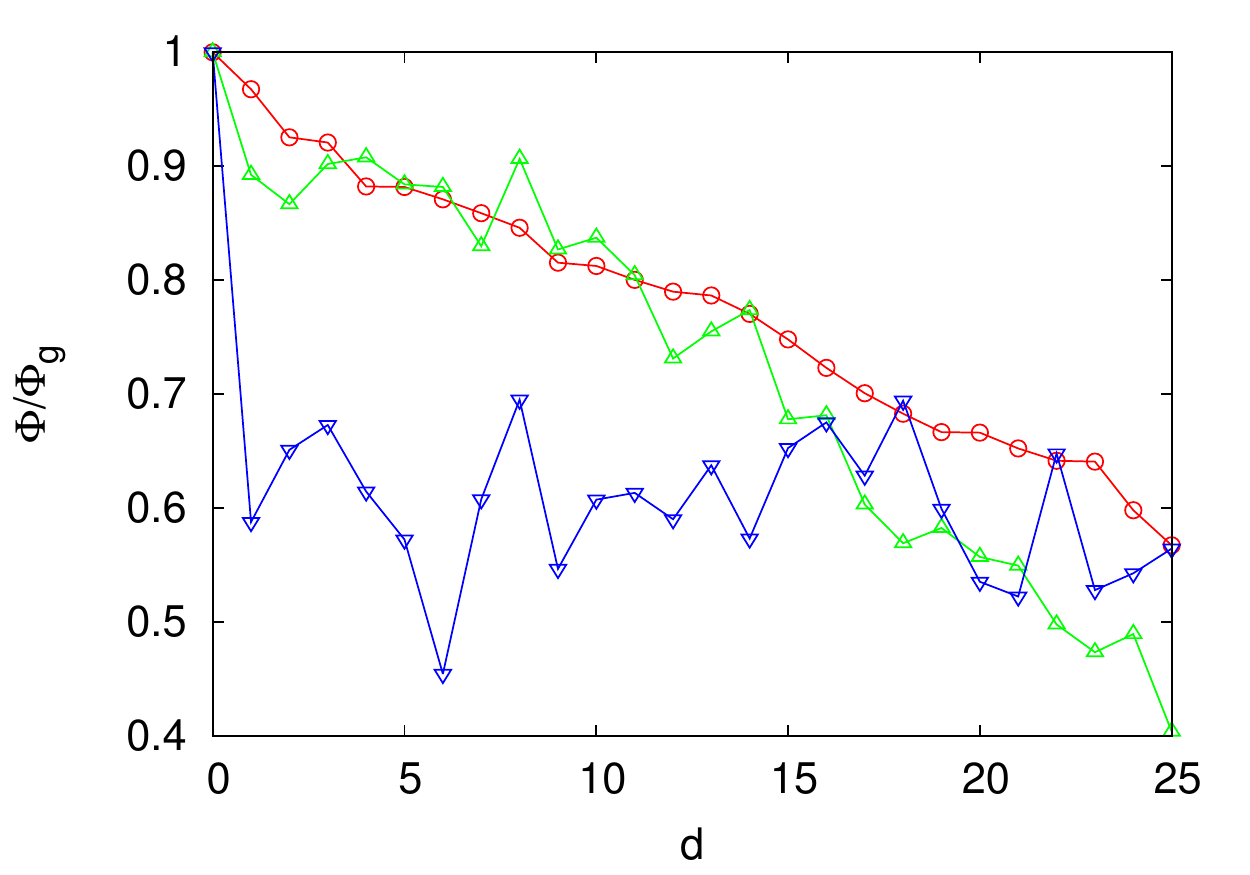}
\caption{(Color online) Normalized fitness as function of the Hamming distance to the global maximum for $N=25$ and $K=0$ (\emph{circles}), $K=5$  (\emph{triangles}) and $K=20$  (\emph{inverted triangles}). The lines are guides to the eye.}
\label{fig:1}
\end{figure}
%
To illustrate the effect of varying $K$ on the ruggedness of the fitness landscape, in Fig.\ \ref{fig:1} we show the  relative fitness
of a string  $\Phi/\Phi_g $ as function of  its Hamming distance $d$ to  the global maximum for $N=25$ and different values of $K$. Here $\Phi_g$ stands for the fitness of the global maximum. For each $K$, the figure shows the results for a single realization of the fitness landscape and for a single trajectory in the solution space that begins at the maximum ($d=0$) and changes the state components sequentially until all $N$ states are reversed ($d=N$). 

We note that   the ruggedness of
the landscapes  (essentially, the number of local maxima)  can vary wildly between landscapes characterized by the same values of $N$ and $K>0$ \cite{Kauffman_87,Kauffman_89} and so can the  performance of any search  heuristic based on the  local correlations of the fitness landscape. Hence in order to appreciate the role of the parameters  that are relevant to imitative learning, namely, the group size $M$ and the imitation probability $p$, for fixed $N$ and $K$ we consider a single realization of the NK fitness landscape.

\section{Imitative Learning Search}\label{sec:Imi}

We assume a group or system  composed of $M$  agents.  Each agent operates in an initial binary string drawn at  random with equal probability for the digits $0$ and $1$. In the typical situation  that the size of the solution space is much greater than the group size
$2^N \gg M$ we can consider those initial strings as distinct binary strings, but here we will consider the case that $M \gg 2^N$ as well, so that
many copies  of a same string are likely to appear in this  initial stage of the simulation. In addition,  we assume that the agents operate in parallel.

At any trial $t$, each agent can choose with a certain probability between two distinct processes to operate  on the strings. The first process,
which happens with probability $1-p$,  is the elementary or minimal move in the solution space, which consists of picking a string bit  $i=1, \dots, N$ at random with equal probability and then flipping it. The repeated application of this
operation  is capable of generating all  the $2^N$  binary strings starting from any arbitrary string.
The second process, which happens with probability $p$, is the  imitation  of a model string. We choose the model string as the highest fitness string in the group at  trial $t$. The string to be updated (target string)  is compared with the model string and the different digits are singled out.  Then the agent  selects at random one of the distinct bits and flips it so that the target string becomes now more similar to the model string. The parameter $p \in \left [0,1 \right ]$ is the imitation probability and the case $p=0$ corresponds to the baseline situation in which  the $M$ agents explore the solution space independently.

The specific imitation procedure proposed here was inspired by
the mechanism used to model the influence of an external media  \cite{Shibanai _01,Avella_10,Peres_11,Peres_12} in  the celebrated agent-based model proposed by  Axelrod to study the process of culture dissemination  \cite{Axelrod_97}.  It is important to note that in the case the target string is identical to the model string, and this situation is not uncommon since the imitation process reduces the diversity of the group, the 
agent executes the elementary move with probability one. This procedure is different from that used in \cite{Fontanari_14}, in which strings identical to the model string are not updated   within the imitation process. However, both procedures yield qualitatively similar results, except in the regime where imitation is extremely frequent, i.e., for $p \approx 1$. In  particular, in the present implementation, a small   group  can find the global maximum in the case  $p=1$ since the model string can explore  the solution space through the elementary move whereas the other strings are simply followers.

The collective search ends when
one of the agents finds the global maximum  and we denote by $t^*$ the number of  trials made by the agent that
found the solution. Of course, $t^*$ stands also for the number of trials made by any one of the $M$ agents, since they operate in
parallel and the search halts simultaneously for all agents. In other words, the trial number $t$ is updated, or more specifically, incremented by one unit, when the $M$ agents have executed  one of the two processes on its associated string. We note that except for the case  $p=0$, the update of the $M$ agents is not strictly a parallel process since the model strings may change several times within a given trial. Nonetheless, since in a single trial all agents are updated, the total number of agent updates at trial $t$ is given by the product $Mt$.

The efficiency of the  search strategy is measured by the total number of agent updates necessary to find the global maximum (i.e., $Mt^*$) which can then be interpreted as  the computational cost of the search.  In addition, 
since we expect that the typical number of trials to success $t^*$ scales with the size of the solution space $2^N$, we will present the results in terms of the rescaled computational cost,  defined as $C \equiv M t^*/2^N$. 

An interesting variant of our imitation rule  is obtained by  relaxing the condition that  only the string with the highest fitness value can be imitated and allowing any string to pose as a  model according to an imitation  probability function proportional to the string relative fitness (e.g., a Fermi function).
Such fitness dependent imitation probability functions are frequently used  in the study of cooperation dilemmas \cite{Perc_10,Wang_14} and may be useful to prevent the  string population being trapped in the local maxima.  In this paper, however,  we will consider only the noiseless or zero-temperature  limit of those imitation functions, where only the string with the maximum relative fitness can be imitated.

\section{Results}\label{sec:res}

Here we report the results of extensive Monte Carlo simulations of  groups of agents that  use imitative learning to  search for
the global maxima of three  representative realizations of the  NK fitness landscape. For fixed $N$ and $K$ we generate a single realization of the fitness landscape and carry out  $10^5$ searches in order to determine the dependence of the  mean rescaled computational cost $ \langle C \rangle $ on the imitation probability $p$ and on the group size $M$. In addition, for the baseline case in which the $M$ agents explore the landscape independently  ($p=0$) we derive an analytical expression for the computational cost.

%
\begin{figure}
\centering
\includegraphics[width=0.48\textwidth]{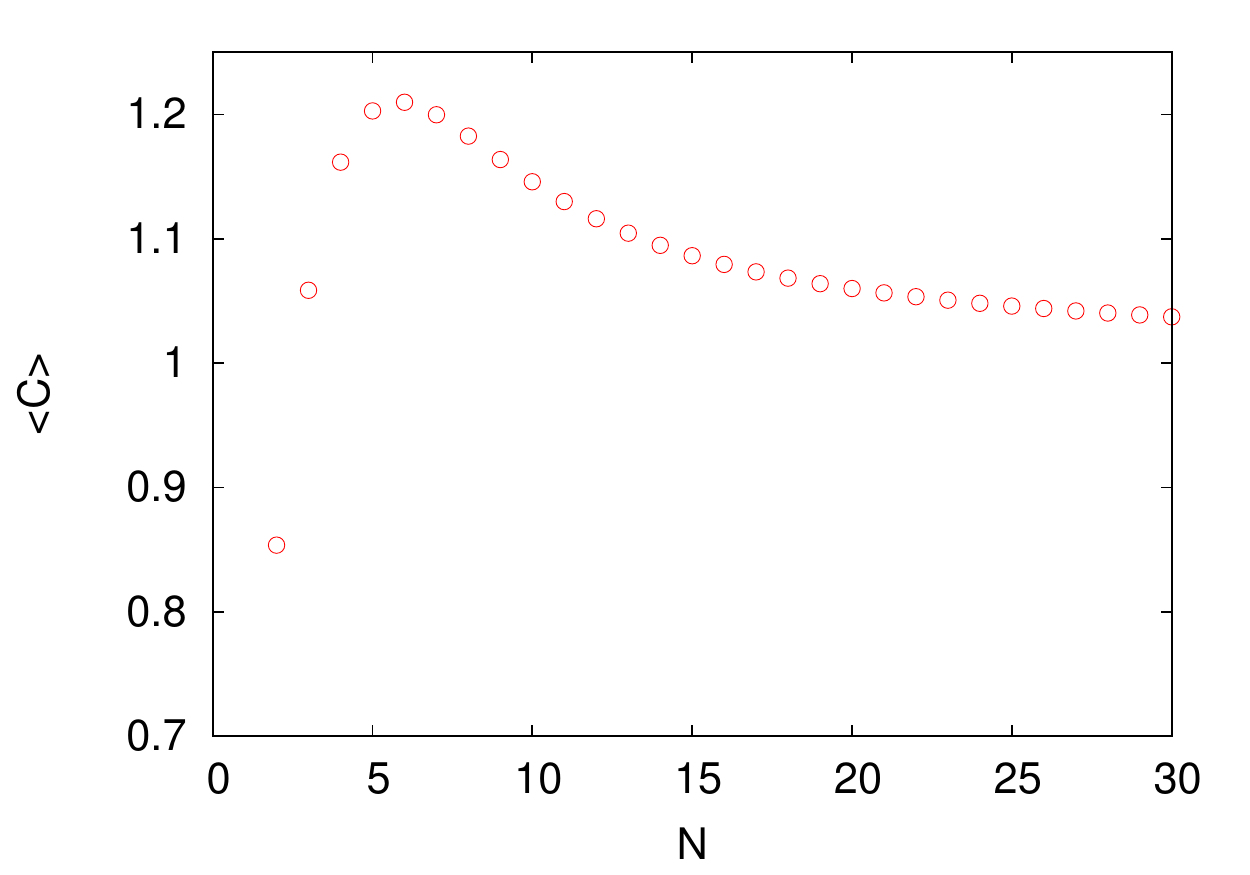}
\caption{(Color online) Single-agent mean rescaled computational cost $\langle C \rangle $ 
as function of the binary string length $N$ that determines the size of the solution space $2^N$. These results, which
were obtained by the numerical calculation of the eigenvalues of the tridiagonal stochastic matrix 
$\mathbf{T}$,  do not depend on the ruggedness of the landscape.
}
\label{new:2}
\end{figure}
%

\subsection{Independent search}\label{sec:Ind}

In this case there is no imitation ($p=0$) and the ruggedness of the landscape has no effect on the performance of the search, which depends only on the size of the solution space, $2^N$. Because of this independence on the landscape we can derive
exact results for the  time needed for $M$ independent agents to find the global maximum. For this analysis we can assume that the global  maximum is the string with  the $N$  digit values equal to  1, i.e., $\left ( 1,1,\ldots, 1 \right) $, without loss of generality. 

Let us consider first the case of a single agent that operates on a string with $j$ digits 1.   According to the elementary move, the probability that the resulting string has $i$ digits 1 is
\begin{equation}\label{Tij}
T_{ij} = \left ( 1-j/N \right ) \delta_{i, j+1} + j/N \delta_{i, j-1}
\end{equation}
for $j=1,\ldots,N-1$, $T_{i0} = \delta_{i,1}$, and  $T_{iN} = \delta_{i,N}$, where $\delta_{i, j} $ is the Kronecker delta.
Since $0 \leq T_{ij} \leq 1$ and $\sum_{i}T_{ij} = 1$ for  $j=0,\ldots,N$ the matrix $\mathbf{T}$ is a tridiagonal stochastic
matrix. The associated stochastic process has a single absorbing state $i=N$ and 
$\pi_N \left ( t \right ) = \sum_{i=0}^{N} T_{Ni} \pi_i \left ( t-1 \right ) $ yields the probability
that the  agent finds the solution  before or at trial $t$.  For large $t$ this quantity is given $\pi_N \left ( t \right ) = 1
- \left( \lambda_N \right )^t$ where $\lambda_N$ is the second largest eigenvalue of $\mathbf{T}$. (The largest eigenvalue
is 1 because $\mathbf{T}$ is a stochastic matrix.) The probability that the agent finds the solution exactly at trial $t=t^*$ is
$p_N \left ( t^* \right )  = \pi_N \left ( t^* \right ) - \pi_N \left ( t^* -1\right )$, i.e.,
\begin{equation}\label{geom1}
p_N \left ( t^* \right )  = \left( \lambda_N \right )^{t^*-1} \left ( 1 - \lambda_N \right ),
\end{equation}
which is a geometric distribution with success probability $1-\lambda_N$. The value of $\lambda_N$ can  be easily obtained
numerically (see, e.g., \cite{Press_92}) provided $N$ is not too large since $2^N \left ( 1 - \lambda_N \right ) $ is on the order of the unity and so it becomes
practically impossible to distinguish $\lambda_N$ from $1$ for $N > 30$.  The mean rescaled computational cost 
$\langle C \rangle = 1/\left [ 2^N \left ( 1 - \lambda_N \right ) \right ]$ is shown in Fig. \ref{new:2} for different string lengths $N$.
(Recall that $M=1$ at this stage.)
In particular for  $N=12$ we find $\lambda_{12} \approx 0.99978$ that implies $\langle C \rangle  \approx 1.11612$. For very large $N$, the probability of success becomes $1 - \lambda_\infty = 1/2^N$. We note that although eq.\ (\ref{geom1}) is valid
strictly for large $t^*$ only, the fact that $t^*$ is typically on the order of $2^N$ makes this geometric distribution an exceedingly good approximation to the correct distribution of absorbing times.

\begin{figure}
\centering
\includegraphics[width=0.48\textwidth]{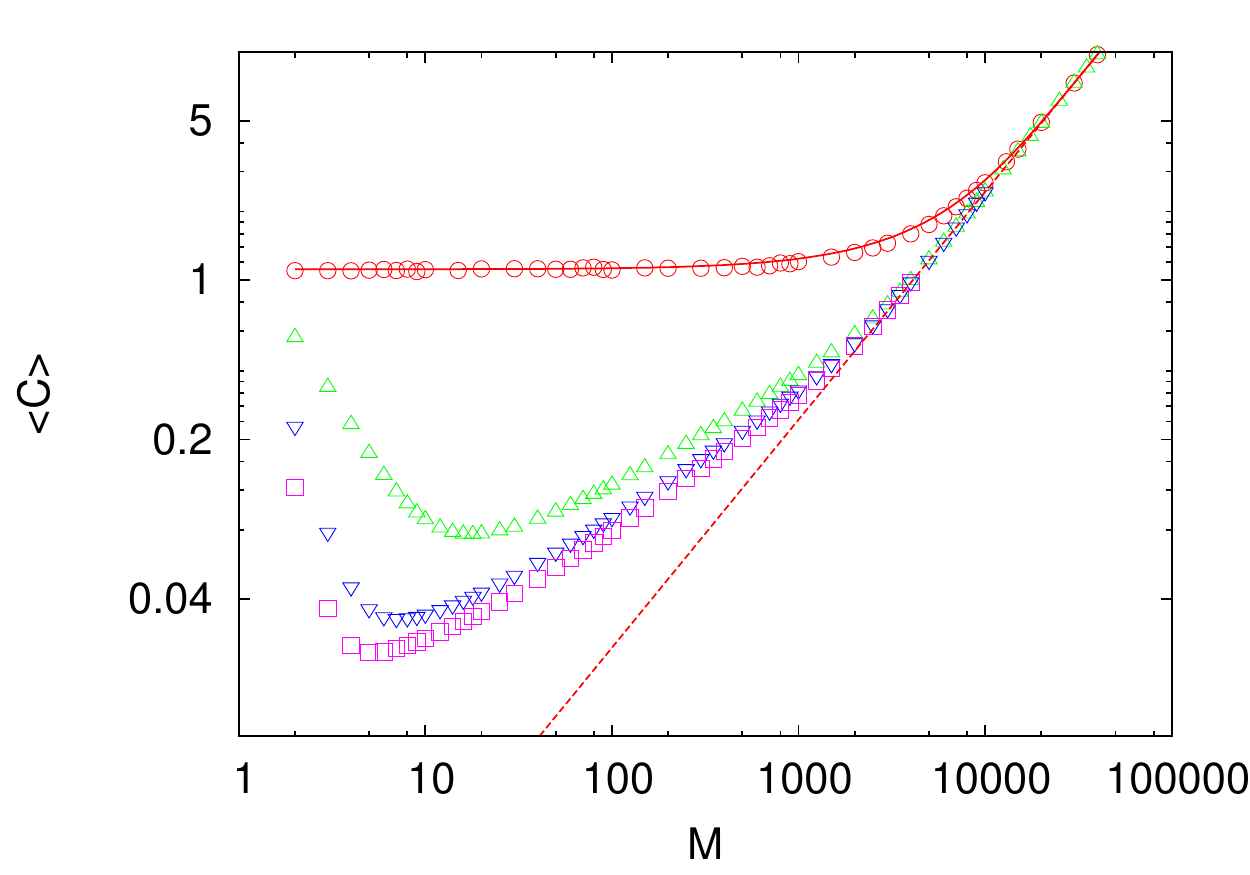}
\caption{(Color online) Mean rescaled computational cost $\langle C \rangle $ as function of the group size $M$ for
the imitation probability $p=0$ (\emph{circles}), $p=0.5$  (\emph{triangles}), $p=0.8$  (\emph{inverted triangles}), and
$p=1$ (\emph{squares}).  The solid  curve is  eq.\ (\ref{Cind}) and the dashed line
is the linear function $\langle C \rangle = M/2^N$. The parameters of the NK landscape are
$N=12$ and $K=0$. The landscape exhibits a single maximum. }
\label{fig:2}
\end{figure}

Now let us consider the case of $M=2$ agents searching independently for the global maximum.  Since the process halts when one of the agents finds the global maximum, the halting time is $t_2^* = \min \left \{ t_a^*, t_b^* \right \} $ where $t_a^*$ and $t_b^*$ are independent random variables distributed by the geometric distribution (\ref{geom1}).  It is easy to show that the distribution of $t_2^*$ is also geometric with probability of success $1 - \left ( \lambda_N \right)^2$ \cite{Feller_68}. In the case of $M=3$ agents the halting time  $t_3^* = \min \left \{ t_2^*, t_a^* \right \} $ and since both $t_2^*$ and $t_a^*$ are geometrically distributed, though with distinct success probabilities, we find that  $t_3^*$ is also geometrically distributed with probability of success $1 - \left ( \lambda_N \right)^3$. The generalization of this reasoning to $M$ agents yields that the mean scaled cost is
\begin{equation}\label{Cind}
\langle C \rangle = \frac{M}{ 2^N \left [ 1 - \left ( \lambda_N \right)^M \right ]}.
\end{equation}
Since $\lambda_N $ is close to $1$ we can write $\left ( \lambda_N \right)^M \approx e^{- M \left ( 1 - \lambda_N \right )}$ so
that $\langle C \rangle \approx 1/\left [ 2^N \left ( 1 - \lambda_N \right ) \right ]$ for $M \ll 1/\left ( 1 - \lambda_N \right )$  and
$\langle C \rangle \approx M/ 2^N $ for $M \gg 1/\left ( 1 - \lambda_N \right )$. We recall that for  $N=12$  we have
$1/\left ( 1 - \lambda_{12} \right ) \approx  4545$. As expected, eq. (\ref{Cind})  matches the simulation data very well (see
Fig.\ \ref{fig:2}).

%
%

\subsection{Search on a Smooth Landscape ($K=0$)}

The  NK fitness landscape with $K=0$  exhibits a single global maximum and no local maxima. The  results of the performance of the imitative search for a landscape with $N=12$ and $K=0$ are summarized in Fig.\ \ref{fig:2}.   

As shown in the previous subsection,  the mean computational cost for non-interacting agents ($p=0$) is a constant $ \langle C \rangle  \approx 1.12 $ provided the group size $M$  is not too large compared to  $1/\left ( 1 - \lambda_{12} \right ) \approx  4545$, which is close to the size of the solution space, $2^{12} = 4096$. 
In the case the group size $M$ is very large, the agents begin to duplicate each other's work  leading to a linear increase of  $ \langle C \rangle$  with  increasing $M$. More pointedly, in this regime we find 
$\langle C \rangle = M/2^N$ (see the straight line in Fig.\ \ref{fig:2}).
We stress that a constant computational cost means that the time $t^*$ the group requires to find the   global maximum decreases with the inverse of the group size (i.e., the time to solve the problem decreases linearly with the number of agents), whereas a computational cost that grows linearly with the group size  means that $t^*$ does not change as more agents are added to the group.

Allowing the agents to imitate the best performer (mo\-del)  at each trial leads to a rapid reduction of the computational cost provided the group size remains small. The best performance is achieved for $M=5$ and $p=1$ and corresponds to a  fiftyfold decrease of the computational cost with respect to the independent search ($p=0$). The fact that the best performance is obtained when the imitation probability is maximum is due to the absence of local maxima in the landscape for $K=0$. We recall that for $p=1$ only the model string, which is likely to be represented by several copies in the group,  can perform the elementary move; all other strings must imitate the model. As a result, the effective size of the search space is greatly reduced, i.e., the strings are concentrated in the vicinity of the model string which cannot accommodate many more than   $M=8$ strings without duplication of work. This is the reason we observe the degradation of the performance when the  group size increases beyond its optimal value. Note that for $K=0$ the imitative learning search  always performs better than the independent search.

\subsection{Search on a Slightly Rugged Landscape ($K=2$)}

Now we consider a somewhat more complex NK fitness landscape by setting $N=12$ and $K=2$. In the particular realization of the landscape
studied here there are 5  maxima in total, among which 4 are local maxima. The relative fitness  of those maxima ($\Phi/\Phi_g$), as well as their Hamming distances  to the global maximum ($d$), is presented in Table \ref{table:k2}.

%
\begin{table}
\caption{Local maxima for the studied realization of the NK fitness landscape with $N=12$ and $K=2$ }
\centering
\begin{tabular}{c c c }
\hline
Maximum  & \hspace{1cm}$\Phi/\Phi_g$ \hspace{1cm} &  $d$ \\ [0.5ex]
\hline
1 & 0.7681 & 5  \\
2 & 0.9114 & 5  \\
3 & 0.9204 & 3   \\
4 & 0.9801 & 3 \\ [1ex]
\hline
\end{tabular}
\label{table:k2}
\end{table}
%

\begin{figure}
\centering
\includegraphics[width=0.48\textwidth]{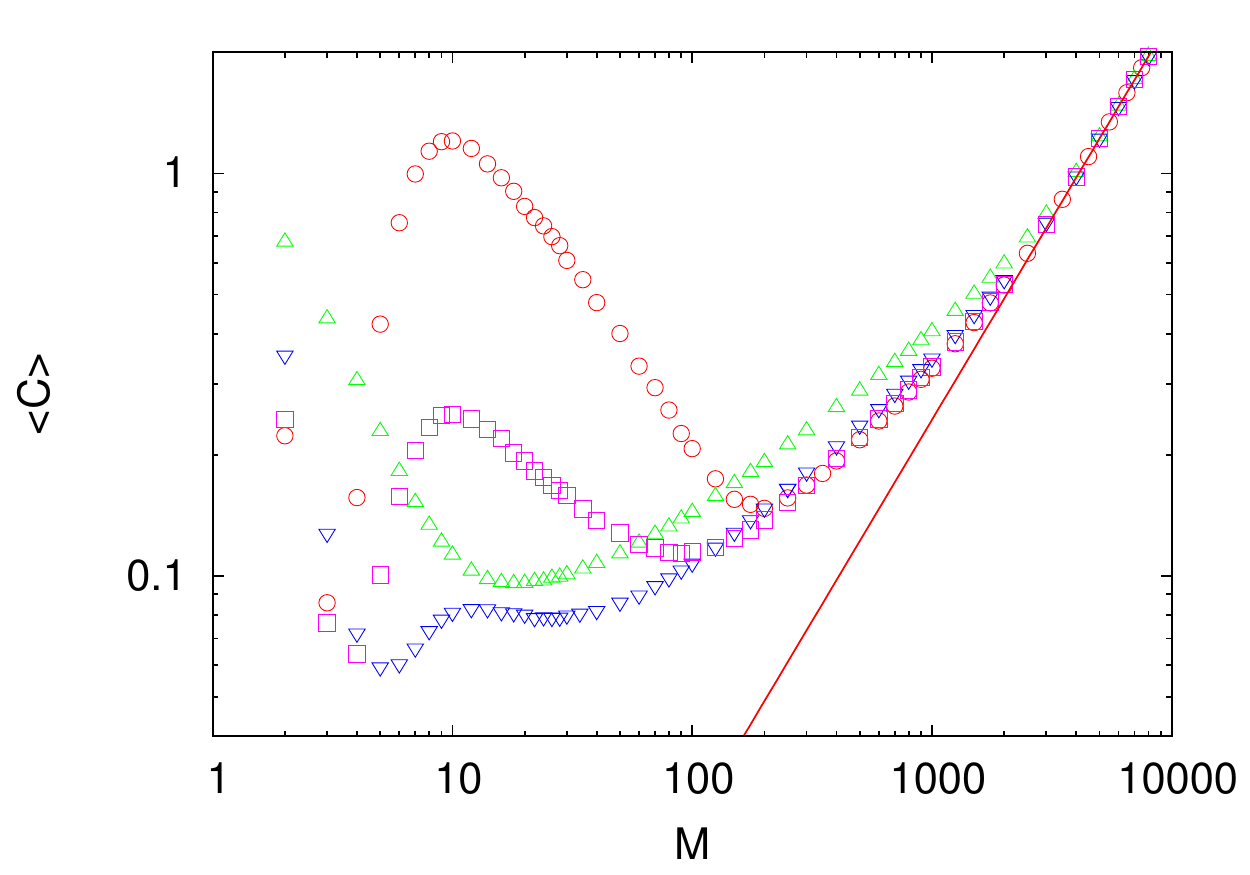}
\caption{(Color online) Mean rescaled computational cost $\langle C \rangle $ as function of the group size $M$ for
the imitation probability $p=0.5$  (\emph{triangles}), $p=0.8$  (\emph{inverted triangles}) , $p=0.95$ (\emph{squares}) and $p=0.99$ (\emph{circles}).  The solid line is the linear function $\langle C \rangle = M/2^N$. The parameters of the NK landscape are
$N=12$ and $K=2$. The landscape exhibits 4 local maxima and a single global maximum.}
\label{fig:3}
\end{figure}

The results of the imitative learning search are summarized in Fig.\ \ref{fig:3} where the mean rescaled computational cost is plotted against the group size for different values of the imitation probability.  The performance of the independent search ($p=0$) is identical to that shown in Fig.\ \ref{fig:2} for $K=0$,
since that search strategy is not affected by the complexity of the landscape. The results for the cooperative search ($p > 0$), however,  reveal a  trade-off between the group size $M$  and the imitation probability $p$. In particular, for $p > 0.8$ we observe a steep  increase of the computational cost for intermediate values of $M$. This happens because  the  group can be trapped near one of the local maxima. For  large groups ($M > 100$ in this case), chances are that some of the initial strings are close to the global maximum and
end up attracting the rest of the group to its neighborhood, thus attenuating the effect of the local maxima.  The robust finding is that for any $p > 0$ there is an optimal group size, which depends on the value of $p$,  that minimizes the computational cost of the search.

%
%

%
\begin{figure}
\centering
\includegraphics[width=0.48\textwidth]{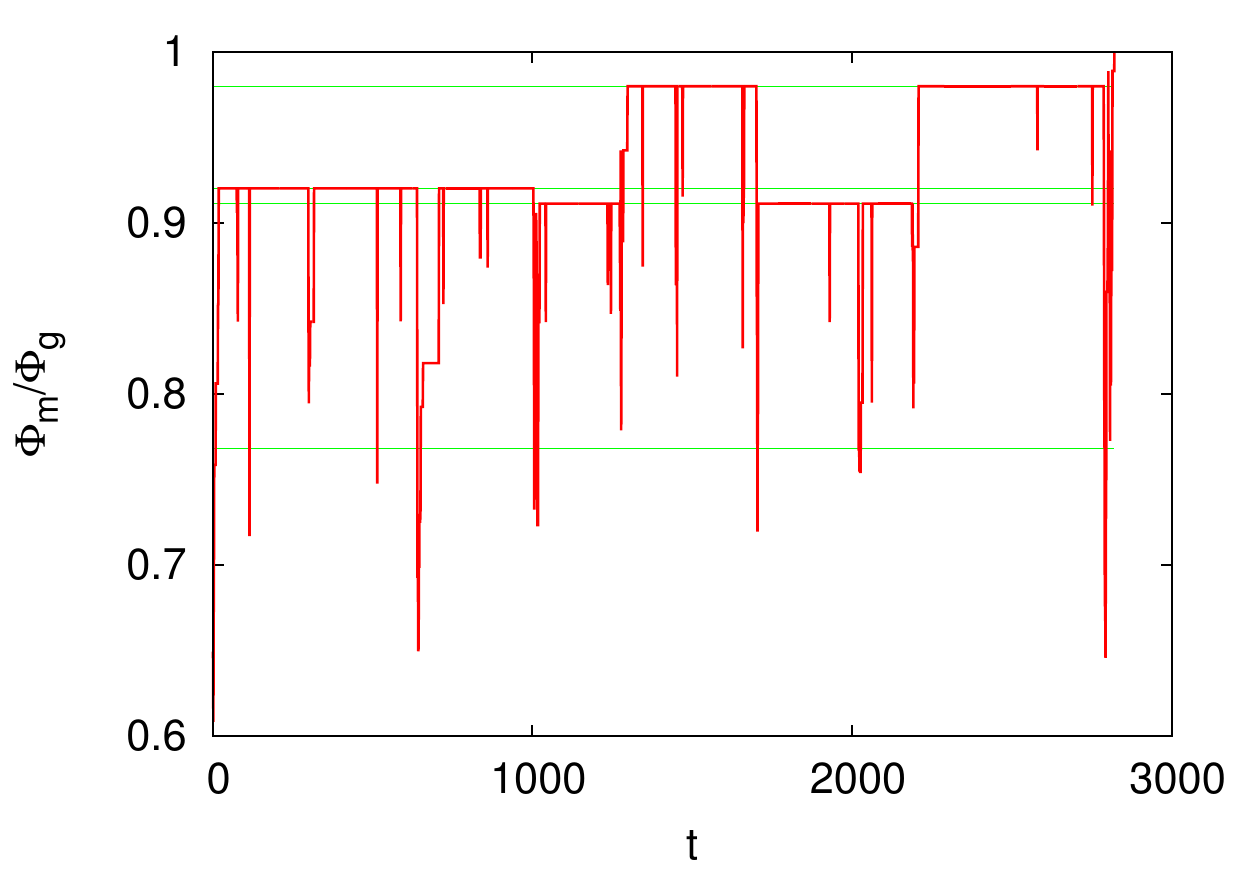}
\caption{(Color online) Time evolution of the relative fitness of the model string for $M=3$ agents and probability of imitation $p=0.99$.   The thin horizontal lines indicate the relative fitness of the  4 local maxima  given in Table \ref{table:k2}. The parameters of the NK landscape are
$N=12$ and $K=2$.}
\label{fig:4}
\end{figure}
%

A better understanding of the dynamics of the search is offered in Fig.\ \ref{fig:4} which shows the relative fitness of the model string as function of the number of trials for $M=3$ and $p=0.99$. The role of the local maxima as transitory attractors of the search is evident in this figure. A typical run with $M=10$ agents, which is approximately the location of the maximum of $\langle C \rangle $ in Fig.\ \ref{fig:3}, yields essentially a flat line at the highest fitness local maximum (maximum 4 in Table \ref{table:k2}) and a sudden jump to the global maximum. Because there are many copies of the model string (the mean Hamming distance between the $M=10$ strings is less than 1), the variants produced by the elementary move  are attracted back to the local maximum. This is the reason why a small group can search the solution space much more efficiently than a large one in the case imitation is very frequent.

\subsection{Search on a Rugged Landscape ($K=4$)}

\begin{figure}
\centering
\includegraphics[width=0.48\textwidth]{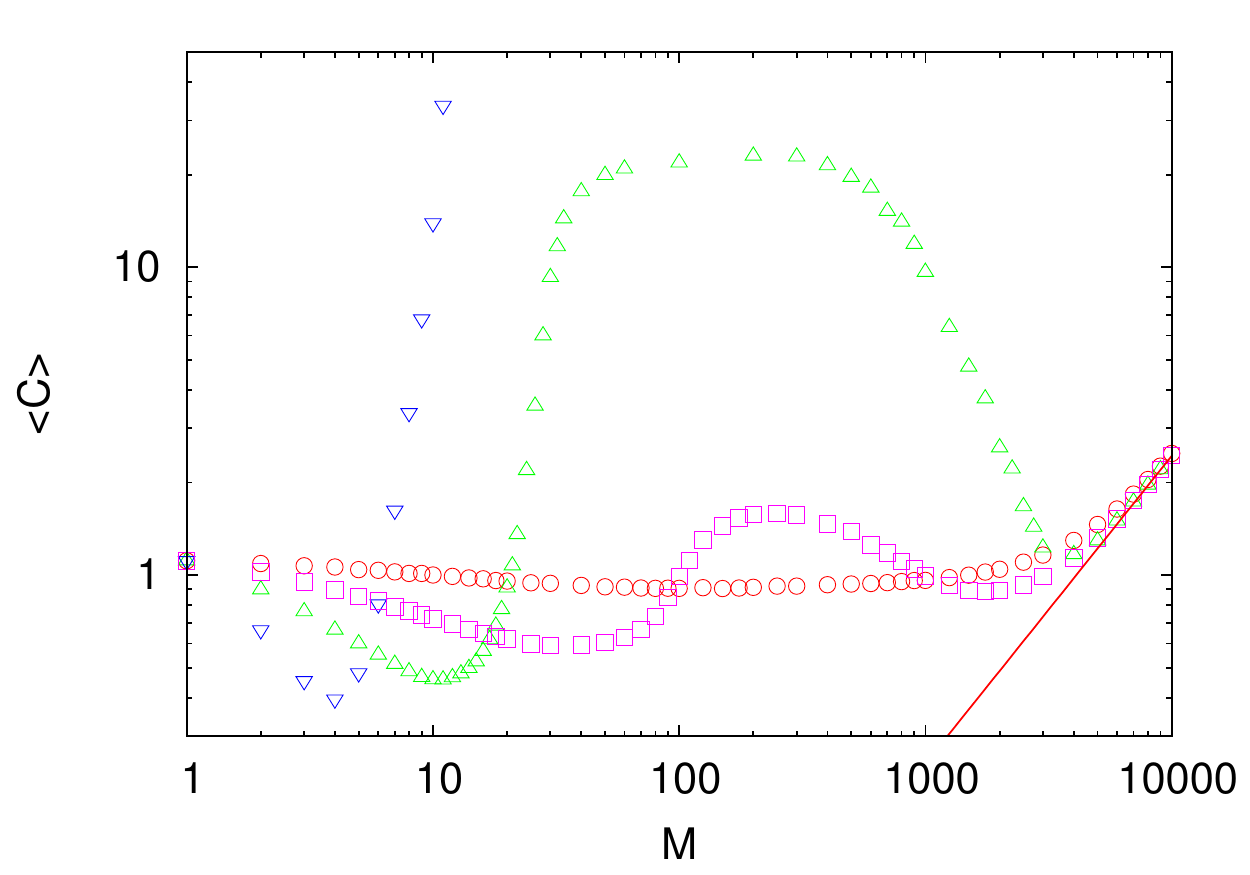}
\caption{(Color online) Mean rescaled computational cost $\langle C \rangle $ as function of the group size $M$ for
the imitation probability  $p=0.1$ (\emph{circles}), $p=0.3$  (\emph{squares}), $p=0.5$  (\emph{triangles}) and  $p=0.8$  (\emph{inverted triangles}).  
For $p=0.8$ we find  $\langle C \rangle > 50$ for $M > 12$. 
The solid line is the linear function $\langle C \rangle = M/2^N$. The parameters of the NK landscape are
$N=12$ and $K=4$. The landscape exhibits 52 local maxima and a single global maximum.}
\label{fig:5}
\end{figure}

The particular realization of the NK fitness landscape with $N=12$ and $K=4$ that we consider now has 53 maxima, including the global maximum, which poses a substantial challenge to any hill-climbing type of search strategy. As in the previous case, we observe in Fig.\ \ref{fig:5} a  trade-off between $M$ and $p$, but now the  results  reveal  how imitative learning may produce  disastrous performances   on rugged landscapes for certain ranges of those parameters.  The strategy of following or imitating a model string can backfire if the fitness landscape exhibits   high fitness local maxima that are distant from the global maximum. A large group may never escape from those
maxima due to  the attractive effect of the clones of the model string. This is what we observed in the case of $p=0.8$ for which we were unable to find the global maximum with groups of size $M > 12$.

%

\begin{figure}
\centering
\includegraphics[width=0.48\textwidth]{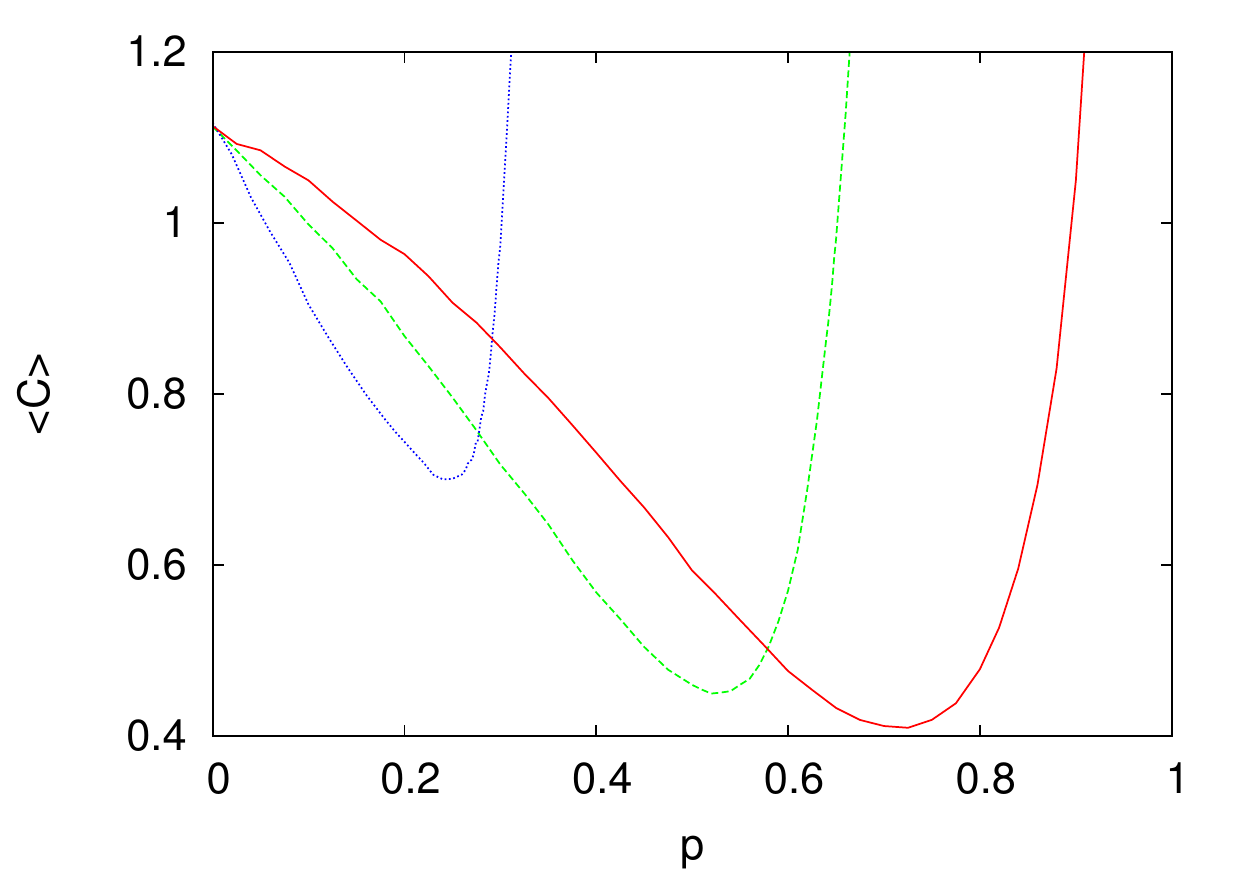}
\caption{(Color online) Mean rescaled computational cost $\langle C \rangle $ as function of the  probability of imitation $p$
for the  group size  $M=5$ (solid line), $M=10$ (dashed line) and $M=100$ (dotted line). The parameters of the NK landscape are
$N=12$ and $K=4$. The landscape exhibits 52 local maxima and a single global maximum.}
\label{fig:6}
\end{figure}
%

We note, however, that for a fixed group size $M$ it is always possible to tune the imitation probability $p$ so that the imitative learning
strategy performs better than (or, in a worst-case scenario, equal to) the independent search.  This point is illustrated in Fig.\ \ref{fig:6} that shows the computational cost as function of $p$. For any fixed value of $M$,  the computational cost exhibits a well-defined minimum that determines the value of the optimal imitation frequency. As hinted in the previous figures, this optimal  value decreases with increasing group size.

\begin{figure}
\centering
\includegraphics[width=0.48\textwidth]{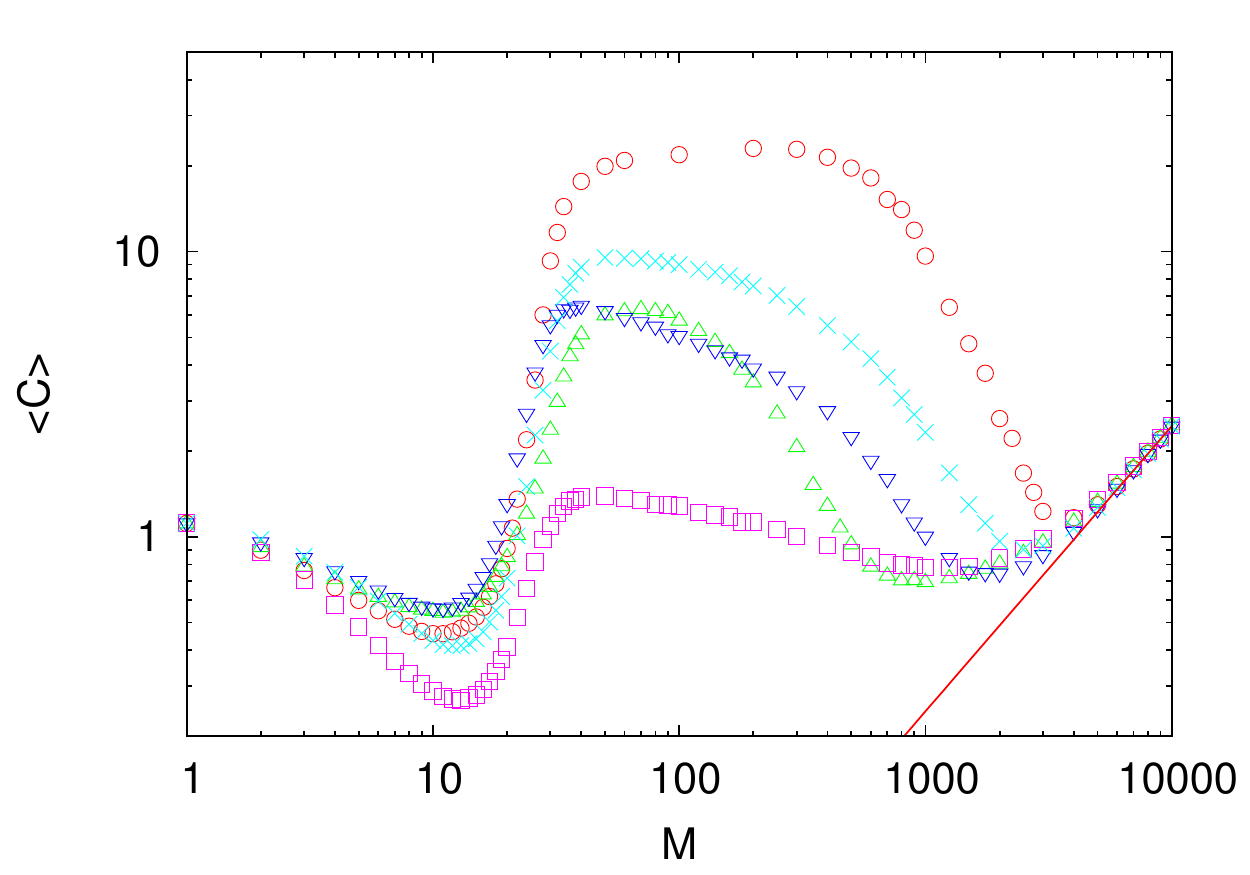}
\caption{(Color online) Mean rescaled computational cost $\langle C \rangle $ for the probability of imitation $p=0.5$ as function of  the  group size $M$  for five realizations (different symbols) of the NK fitness landscape  with $N=12$ and $K=4$. The solid line is the linear function $\langle C \rangle = M/2^N$. }
\label{fig:8}
\end{figure}

In order to verify the robustness of our findings, which were obtained for specific realizations of the NK fitness landscape, we have considered  four random realizations of the landscape with $N=12$ and $K=4$ in addition to the realization studied above.  The comparison between the mean computational costs to find the global maxima of  the five realizations of the landscape  is shown in Fig.\ \ref{fig:8}  for the imitation probability $p=0.5$. The results are qualitatively the same, despite  the wild
fluctuations of $\langle C \rangle $  in the regime where the search is trapped in the local maxima.   It is reassuring to note
that the initial decrease of  the mean cost with increasing group size and the existence of an optimal group size that
minimizes  that cost, which are exhibited by all five realizations of the  NK landscapes shown in Fig.\ \ref{fig:8}, are robust  properties of the imitative learning search.

%

\section{Discussion}\label{sec:conc}

In this paper we  study quantitatively the problem solving performance  of a group of   agents capable to
learn by imitation.  The performance measure we consider  is  the computational cost  to  find the global maximum of three specific realizations of the NK fitness landscape. The computational cost is defined as  the product
of  the number of agents in the group and the  number of  attempted  trials till some agent finds the global maximum. Our main conclusion,
namely, that for a fixed probability of imitation $p$ there is a value of group size  that minimizes the computational cost corroborates the findings
of a similar study in which the task was to solve a particular cryptarithmetic problem \cite{Fontanari_14}. Hence our conjecture that 
the efficacy of imitative learning could be a factor determinant of the group size of social animals (see \cite{Wilson_75,Dunbar_92} for a discussion on the  standard selective pressures on group size in nature). We note that in the case the connectivity between agents is variable,
i.e., each agent interacts with $L=1,\ldots,M-1$ distinct randomly picked agent (here we have focused on the fully connected network $L=M-1$ only)  then there is an optimal connectivity value that minimizes the computational cost \cite{Fontanari_14}. It would be most interesting to understand how the network topology influences the performance of the group of imitative agents and how the optimal network topology  correlates with the known animal social networks \cite{Wey_08}.

The main aim  of our contribution is to show that  the  existence of an optimal group size that maximizes performance for imitative learning 
\cite{Fontanari_14} is insensitive to the choice of the fitness landscape, so it is likely a robust property of populations that use imitation as a cooperative strategy. Although we have focused on the effect of the parameter $K$, which roughly determines the ruggedness of the NK  landscape, the parameter $N$ (the length of the strings) also plays  a relevant role on the search for the global maximum, besides the obvious role of fixing the size $2^N$ of the search space. (Of course, since the typical time to find the global maximum scales with $2^N$, even moderate   values of $N$, say $N=16$,  would render the simulations unfeasible.)  The nontrivial role is that the value of $N$  seems to pose an upper bound to the  optimal size of the group $M$ in the regime that  imitation is very frequent (see Figs. \ref{fig:2}, \ref{fig:3} and \ref{fig:5}). This is so because in that regime  the strings are concentrated in the close vicinity of the model string, which cannot accommodate more than  $M=N$ different strings.

We do not purport to offer here any  novel  method to explore efficiently rugged landscapes, but the finding that for small group sizes  imitative learning  decreases considerably  the computational cost of the search, even in a very rugged landscape (see data for $p=0.8$ in Fig.\ \ref{fig:5}) motivates us to address the question whether in such landscapes that strategy could achieve a better-than-random  performance for all group sizes. This  is  achieved automatically for smooth landscapes (see Figs.\ \ref{fig:2} and
\ref{fig:3}) but not for rugged ones (see Fig.\ \ref{fig:5} and \cite{Fontanari_14}).  Clearly, the way to accomplish this aim  is to decrease the probability of imitation $p$ as the group size $M$ increases, following the location of the minima shown in Fig.\ \ref{fig:6}. It is interesting to note that the finding that too frequent  interactions between agents may harm the performance of the group (see Fig.\ \ref{fig:6}) may offer a theoretical justification for  Henry Ford's factory design in which the communication between workers was minimized in order to maintain   the established efficiency and maximal productivity \cite{Watts_06}.

\acknowledgments
This research was partially supported by grant
2013/17131-0, S\~ao Paulo Research Foundation
(FAPESP) and by grant 303979/2013-5, Conselho Nacional de Desenvolvimento 
Cient\'{\i}fico e Tecnol\'ogico (CNPq).

\end{document}